\documentclass[useAMS,usenatbib]{mn2e}

\usepackage{amsmath,texlogos}
\usepackage{natbib}
\usepackage{subfigure}
\usepackage{graphicx}
\usepackage{txfonts}
\usepackage[T1]{fontenc}
\usepackage{aecompl}
\usepackage{times}

\title[The SIDRA algorithm]{SIDRA : a blind algorithm for signal detection in photometric surveys.}
\author[D. Mislis]{D. Mislis$^{1}$ \thanks{E-mail: dmislis@qf.org.qa}, E. Bachelet$^{1}$, K. A. Alsubai$^{1}$, D. M. Bramich$^{1}$, N. Parley$^{1}$ \\
$^{1}$Qatar Environment and Energy Research Institute (QEERI), HBKU, Qatar Foundation, PO Box 5825, Doha, Qatar\\
}
 
\begin{document}

\date{Accepted . Received 1; }

\pagerange{\pageref{firstpage}--\pageref{lastpage}} \pubyear{2014}

\maketitle

\label{firstpage}

\begin{abstract}
We present the Signal Detection using Random-Forest Algorithm (\texttt{SIDRA}). \texttt{SIDRA} is a detection and classification algorithm based on the Machine Learning technique (Random
Forest). The goal of this paper is to show the power of \texttt{SIDRA} for quick and accurate signal detection and classification. We first diagnose the power of the method 
with simulated light curves and try it on a subset of the \textit{Kepler} space mission catalogue. We use five classes of simulated light curves (CONSTANT, TRANSIT, VARIABLE, MLENS and EB for constant light 
curves, transiting exoplanet, variable, microlensing events and eclipsing binaries, respectively) to analyse the power of the method. The algorithm uses four \textit{features} in 
order to classify the light curves. The training sample contains 5000 light curves (1000 from each class) and 50000 random light curves for testing. The total \texttt{SIDRA} 
success ratio is $\geq 90\%$. Furthermore, the success ratio reaches 95 - 100\% for the CONSTANT, VARIABLE, EB, and MLENS classes and 92\% for the TRANSIT class with a decision probability 
of 60\%. Because the TRANSIT class is the one which fails the most, we run a simultaneous fit using \texttt{SIDRA} and a Box Least Square (BLS) based algorithm for searching for 
transiting exoplanets. As a result, our algorithm detects 7.5\% more planets than a classic BLS algorithm, with better results for lower signal-to-noise light curves. 
\texttt{SIDRA} succeeds to catch 98\% of the planet candidates in the \textit{Kepler} sample and fails for 7\% of the false alarms subset. \texttt{SIDRA} promises to be useful for developing 
a detection algorithm and/or classifier for large photometric surveys such as TESS and PLATO exoplanet future space missions.
\end{abstract}



\section{Introduction}

The recent development of wide field photometric surveys opens up a new field of astrophysics. The deployment of both ground based (SuperWASP, HAT, QES) and space based 
(\textit{CoRoT},\textit{Kepler}) surveys increases dramatically our knowledge about transiting planets. Indeed, the huge amount of collected data leads to a real problem of 
identifying targets. The OGLE survey, for example, observes more than 300 million stars in the Galactic bulge each night, leading to a sorting problem. This kind of problem is 
a known as a Big Data problem.\\
Several methods are used to tackle the Big Data problem, and the Machine Learning algorithm is one of them. The Random Tree/Forest algorithm was described in 2001 by L. 
Breiman \citep{Breiman} as part of Artificial Intelligence and Machine Learning general algorithms. Some teams have already used Machine Learning algorithms for astronomical 
projects, especially for the large amount of data from the \textit{Kepler} mission \citep{KeplerML}. Machine Learning object detection and classification for automated classification 
of active stars and galaxies is described in \cite{Li}, using the k-Nearest Neighbours method. Recently, \cite{masci} published an algorithm based on Random Forest for automatic 
classification of variable stars using the \textit{Wide-field Infrared Survey Explorer} data with a success ratio from 87.8 to 96.2\%. In the same year, OGLE detected a supernova 
Type Ia event, using real-time detection and Machine Learning automatic classification \citep{wyrzykowski}. Furthermore, a Random Forest algorithm is used by \cite{mccauliff} to 
identify false positives in the \textit{Kepler} mission data.\\
The exoplanet microlensing surveys such as OGLE and MOA are facing a challenge with real-time photometry and lens detection \citep{bond}. Microlensing detections must 
be observed by as many teams as possible in order to have a complete phase coverage of the phenomenon. This introduces a need for fast event detection on a huge amount of light 
curves.\\
In 2002, \cite{kovacs} published the Box Least Square (BLS) algorithm for transiting exoplanet detection and since then, there are many different versions of BLS \citep{Foreman-Mackey}. 
BLS is a very successful algorithm for almost all of the transiting surveys such as \textit{Kepler} \citep{boruki}, \textit{CoRoT} \citep{moutou}  \textendash from space  \textendash and SuperWASP 
\citep{cameron}, HATNet \citep{bakos}, QES \citep{alsubai} etc. \textendash from the ground. In principal, in order to detect a transiting signal in a light curve using a BLS algorithm, we have to fit the 
orbital period of the planet, the centre of the eclipse $T_{\rm C}$, the duration of the transit and the depth of the transit \citep{borde, bonomo, cabrera}. \\
In this paper, we study a very different approach for signal detection and classification for transiting exoplanets, variable stars and microlensing events by changing the 
philosophy of signal detection from fitting to blind search using Machine Learning techniques. \\

\section{Description of the method and simulated light curves} \label{sec:dis}

\subsection{Light\textendash curve simulations} \label{sec:lcsim}
\begin{figure}
\centering
\includegraphics[width=9cm]{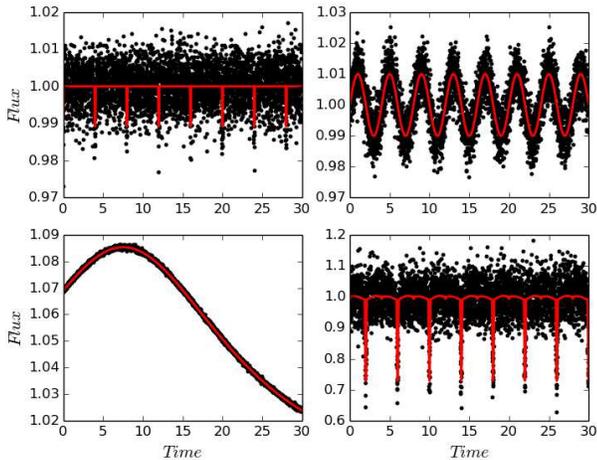}
\caption{Four random light curves (TRANSIT, VARIABLE, MLENS and EB) and their model (red solid line).}
\label{fig1}
\end{figure}

In this paper, we used simulated light curves for the training and testing sample in order to perform various tests. We focused on five typical light curve types which can be 
expected in photometric surveys. These are constant stars (called hereafter CONSTANT), the exoplanet transiting light curves (TRANSIT), the variable stars (VARIABLE), the eclipsing 
binaries (EB) and the microlensing light curves (MLENS). Each light curve is described by a normalized flux as a function of time. We added noise to each light curve with various 
precisions. The rms is selected from a uniform distribution between 1 and 5\%. We used a 30-d observing window, with a 30 min sample to simulate the light curves. Some of these 
types of light curves, such as the TRANSIT sample, require stellar physical properties (stellar mass, stellar radius, effective temperature) given by \cite{kaler}. We select main 
sequence host stars randomly, using a uniform distribution from F0 to M5 spectral type. This range was adopted because it roughly represents 90\% of the total stars in the 
sky \citep{robin}. Table \ref{tab:t1}, summarizes all the input parameters and ranges we used for the simulated light curves. Fig. \ref{fig1} shows typical simulated light curves 
from each class. A more detailed description for each light curve class is given below.

\subsubsection{CONSTANT} \label{sec:const}

This subset of light curves is the most simple, and we can easily create it using pure white noise. The flux $f_{\rm C}$ of a constant light curve is given by 
\begin{equation}
f_{\rm C}(t)=1+\epsilon 
\end{equation}
\noindent
where $\epsilon$ is a random variable set by a normal distribution $N(0,$rms$)$. The rms is randomly selected to be in the range of $0.01 \leqslant $ rms $ \leqslant 0.05$. 

\subsubsection{TRANSIT} \label{sec:tra}
The host stars spectral type and the physical properties are selected randomly from the main-sequence data set as we describe in Section \ref{sec:lcsim}. The planet radius is 
also chosen randomly to be in the range 0.7 - 2 $R_{\rm J}$ and the period was chosen between 1 and 15d. The inclination angle $i$ was chosen in the range 
$i_{\rm min} \leqslant i \leqslant 90$ deg to ensure a transit. The $i_{\rm min}$ is the minimum transit angle described by
\begin{equation}
\cos{i_{\rm min}}={{R_{\star}+R_{\rm P}}\over{\alpha}}
\end{equation}
where $R_{\star}$ is the star radius, $R_{\rm P}$ the planetary radius and $\alpha$ the semi\textendash major axis. Note that we choose a null eccentricity. The transit model is produced using 
a quadratic limb-darkening law and the adopted flux is given by 
\begin{equation}
f_{\rm T}(t)=Pal(t,P,\delta,i,d)+\epsilon
\end{equation}
where Pal$()$ is the \cite{pal} transiting analytical model, $P$ is the period, $\delta$ is the transit depth and $d$ is the duration of a transit. The limb\textendash darkening coefficients 
are given by \cite{claret}. 

\subsubsection{VARIABLES} \label{sec:var}
For this subset, we selected only stellar types crossing the main sequence and the instability strip of the Hertzsprung\textendash Russel diagram. This leads to select only F\textendash type stars.
We modelled the light curve by using a pure sinusoidal signal, with a flux amplitude $0.01 \leqslant A_{\rm V} \leqslant 0.1$ and period $P$, between 1 and 15d. This way we 
mainly modelled some common types of variable stars such as RR Lyrae, $\delta$-Scuti etc. Equation \ref{eq:var} gives the flux of our variable subset light curves : 
\begin{equation}
f_{\rm V}(t)=1+A_{\rm V} \cdot sin(2\pi t/P)+\epsilon.
\label{eq:var}
\end{equation}

\subsubsection{Eclipsing binaries} \label{sec:data}
The eclipsing binaries subset light curves were modelled using the same stellar main sequence characteristics (Section \ref{sec:lcsim}). The selected period is again between 1 and 
15d. In order to simulate a full eclipsing binary light curve (primary/secondary eclipse, ellipsoidal variations etc) we used the PHOEBE eclipsing binary analytical model 
\citep{prsa}. PHOEBE requires stellar characteristics for both of the stars, such as stellar temperatures, masses and radii plus the orbital period of the system. We do not describe 
here the full equation package but the light\textendash curve flux $f_{\rm EB}$ we used is given by equation \ref{eq:bin} :
\begin{equation}
f_{\rm EB}(t) = PHOEBE(t,M_{i}, R_{i}, T_{i},P)+\epsilon ,
\label{eq:bin}
\end{equation}
where PHOEBE$()$ is the eclipsing binaries model, $M_{ i}$, $R_{ i}$ and $T_{ i}$ are the stellar mass, radius and temperature respectively for the primary 
$(i=1)$ and the secondary $(i=2)$ star. Finally, $P$ is the orbital period. 

\subsubsection{Microlensing} \label{sec:mlens}
We produce microlensing light curves as though they have been observed by a survey such as MOA \citep{bond} or OGLE \citep{udalski}. We select a uniform random value for the time 
of maximum $t_{\rm O}$ between 0 and 30d. We also select $U_{\rm o}$, the minimum impact parameter, from a uniform distribution \citep{alcol, sumi}. We select the Einstein ring 
crossing time $t_{\rm E}$ from a normal distribution with a mean of 20d and a standard deviation of 5d, which is a rough approximation of the true $t_{\rm E}$ 
distribution \citep{sumi}. Finally, we select a uniform distribution for the source flux $f_{\rm s}$ and the blend flux $f_{\rm b}$ in the range of 1\textendash 	10.
The adopted flux is
\begin{equation}
f_{M}(t)={{f_{\rm s} \cdot A(t)+f_{\rm b}}\over{f_{\rm s}+f_{\rm b}}}+\epsilon
\end{equation}
where $A(t)$ is the microlensing magnification \citep{paczynski}.

\noindent

\begin{table}
\centering
\caption{\label{tab:t1} Input parameters and limits for our simulated light\textendash curves sample. For each value we use uniform 
random distribution.}
\begin{tabular}{lc}
\hline
rms $(\epsilon)$ & 0.01\textendash 0.05    \\
Period $(P)$ & 1\textendash 15d      \\
Spectra type & F0\textendash M5   \\
Planetary radius $(R_{\rm P})$ & 0.7\textendash 2.0 $R_{\rm J}$   \\ 
Transit inclination $(i)$ & $i_{\rm min}$\textendash 90 deg  \\
Observing window $(t)$ & 30d \\ 
Time resolution & 30 min \\ 
\hline
\end{tabular}
\end{table}

\subsection{Random Forest basics} \label{sec:data}
In the Machine Learning domain, the key is to give informative parameters to the algorithm that describe the problem in hand. These parameters, called features, must 
reflect the intrinsic properties of the different classes. Using these features as inputs, the Random Forest is a three step algorithm, as a typical Machine Learning procedure 
suggests (train-test-predict).
\begin{itemize}
  \item The first step is the training part of the algorithm. The Random Forest uses the features of each vector of the training sample to build $N_{\rm tree}$ decision trees which 
  are tuned to fit the output classes (train-step). 
  \item After the training process, it is recommended to characterize the performance of the algorithm by using an exercise sample (test-step).	 
  \item If the user is satisfied with the accuracy of Test-Step, then the Random Forest can be used for any feature of the sample (predict-step). 
\end{itemize}
To help in the customization of the Random Forest algorithm, we used various tools. The \textit{feature importance} vector gives the relative importance of each feature to produce 
the most accurate estimator. The \textit{confusion matrix} shows in a simple way how well the algorithm performs. Its diagonal values are equal to the \textit{success ratio} for each 
of the classes. Also, the $i \neq j$ element of the \textit{confusion matrix} give the false positive/negative rates \citep{masci}.
\subsection{The statistical method} \label{sec:stats}
\texttt{SIDRA}, basically follows three simple rules :
\begin{itemize}
  \item Features must be as general as possible. \texttt{SIDRA} is able to compute them in a fully blind way for all kinds of light curves.
  \item Features extraction must also be as fast as possible, in order to make the algorithm useful for large and/or real-time surveys (\textit{TESS}, \textit{PLATO}, OGLE, ATLAS etc).	 
  \item Features must show very weak correlation with each other. There is no additional information for highly correlated features. 
\end{itemize}
To respect the first and second conditions, we make the choice to derive our features without any model fits and we calculate only the \textit{statistical information} straight from 
the light curve. We use four features for \texttt{SIDRA}.
\begin{itemize}
  \item The skewness $S$ : this is a measure of the asymmetry of a distribution, defined as the third standardized moment
\begin{equation}
S={{1}\over{n}}\sum\limits_{i=1}^n{{(x_{i}-\bar{x})^3}\over{\sigma^3}} ,
\label{skew}
\end{equation}
where $\bar{x}$ is the mean, $\sigma$ the standard deviation and $n$ the total number of observations. \newline

  \item The kurtosis $K$ : this is a measure of the flatness of a distribution defined as the fourth standardized moment :
\begin{equation}
K=\sum\limits_{i=1}^n{{(x_{i}-\bar{x})^4}\over{\sigma^4}}
\label{kurt}
\end{equation}	
 \newline
  \item The autocorrelation integral $A_{\rm I}$ : the autocorrelation integral is the sum of the autocorrelation values for all possible delays $\tau$. The autocorrelation versus 
  delay $(\tau)$ diagram gives information about periodical patterns of the light curve. For \texttt{SIDRA}, we explore the full observation window and measure the integral given 
  by the autocorrelation vector
\begin{equation}
A_{\rm I} = \left| \sum\limits_{\tau=1}^{n}{ \left(  \frac{1}{(n-\tau)\cdot rms^{2}}     \sum\limits_{i=1}^{n=\tau}{ \left( x_{ i} - \bar{x} \right) \left( x_{ i+\tau} - \bar{x} \right)  } \right) } \right|
\label{ai}
\end{equation}
\newline
  \item The modified information entropy $(E_{\rm S})$ - or Shannon Entropy \citep{shannon} : for each class of light curve, we assume a normal distribution. This is true only for a 
  CONSTANT light curve, but still there is more information for all the other light curve types too. Thus, each $x_{i}$ has a probability based on the Cumulative Distribution 
  Function (CDF). Based on the nature of the survey (exoplanets, variables, microlensing), we can use different CDFs (normal or inversed Gaussian CDF \textendash Fig. \ref{entro}) 
  in order to cover different light curve cases. For our current \texttt{SIDRA} version we used the normal and the inversed Gaussian CDF (blue and red line \textendash 
  Fig. \ref{entro}) in combination.
\begin{figure}
\centering
\includegraphics[width=9cm]{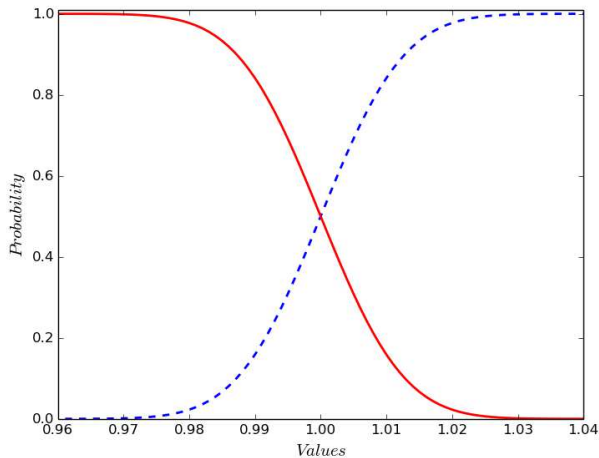}
\caption{Two different probability functions. A normal Gaussian CDF (blue dashed line) and the inversed Gaussian CDF (red solid line).}
\label{entro}
\end{figure}
\noindent
The probability $P(x_{i})$ of each data point $x_{i}$ is given by equation \ref{prob}:
\begin{equation}
P(x_{i}) = \frac{1}{2} \cdot \left( 1+\frac{2}{\sqrt{\pi}} \int_{0}^{ \frac{x_{i}-\bar{x}}{rms \cdot \sqrt{2}} } e^{-t^{2}}  dt  \right) ,
\label{prob}
\end{equation}
\noindent
and the total entropy $E_{\rm S}$ of each light curve is given by equation \ref{ent}
\begin{equation}
E_{S}(x) = -\sum_{i=1}^{i=n} \left( \int_{ \delta_{1}}^{\delta_{2}} \log_{2}(P(\delta))d\delta \right) ,
\label{ent}
\end{equation}
\noindent
where $\delta_{1,2}=x_{i} \pm \sigma_{i}$ and $\sigma_{i}$ is the error of the point $x_{i}$ of the light curve.  
\end{itemize}
\noindent
Our final $E_{\rm S}$ is calculated by adding two values of $E_{\rm S}$ calculated by the normal and inverse Gaussian CDF.\\
The features we have chosen show weak correlation in the parameter space. Fig. \ref{fig4} shows the correlation matrix between all features and classes. VARIABLE, EB and MLENS 
classes are very well determined. On the other hand there is a confusion between CONSTANT and TRANSIT classes. Because of the low signal-to-noise (S/N) ratio of some light curves, 
it is impossible to distinguish between constant and transits (pure noise light curves). Fig. \ref{fig4} \textendash (bottom), explains the results of the \textit{confusion matrix} 
(Fig. \ref{fig3}).\\ 
It is clear from Fig. \ref{fig4} \textendash (top) that in some cases the Random Forest decision is very obvious because the classes are very well separated, suggesting that 
maybe we do not need a Random Forest algorithm. On the other hand, Random Forest becomes important to distinguish objects where their features are mixed (Fig. 
\ref{fig4} \textendash bottom). Also, in this paper we show only some cases. The input classes could be modified by any team, or even increased by adding more different light curves, 
making the problem much more complicated (distinguish between supernova \textendash microlensing light curves or different types of variables for example).
\begin{figure}
\centering
\includegraphics[width=9cm]{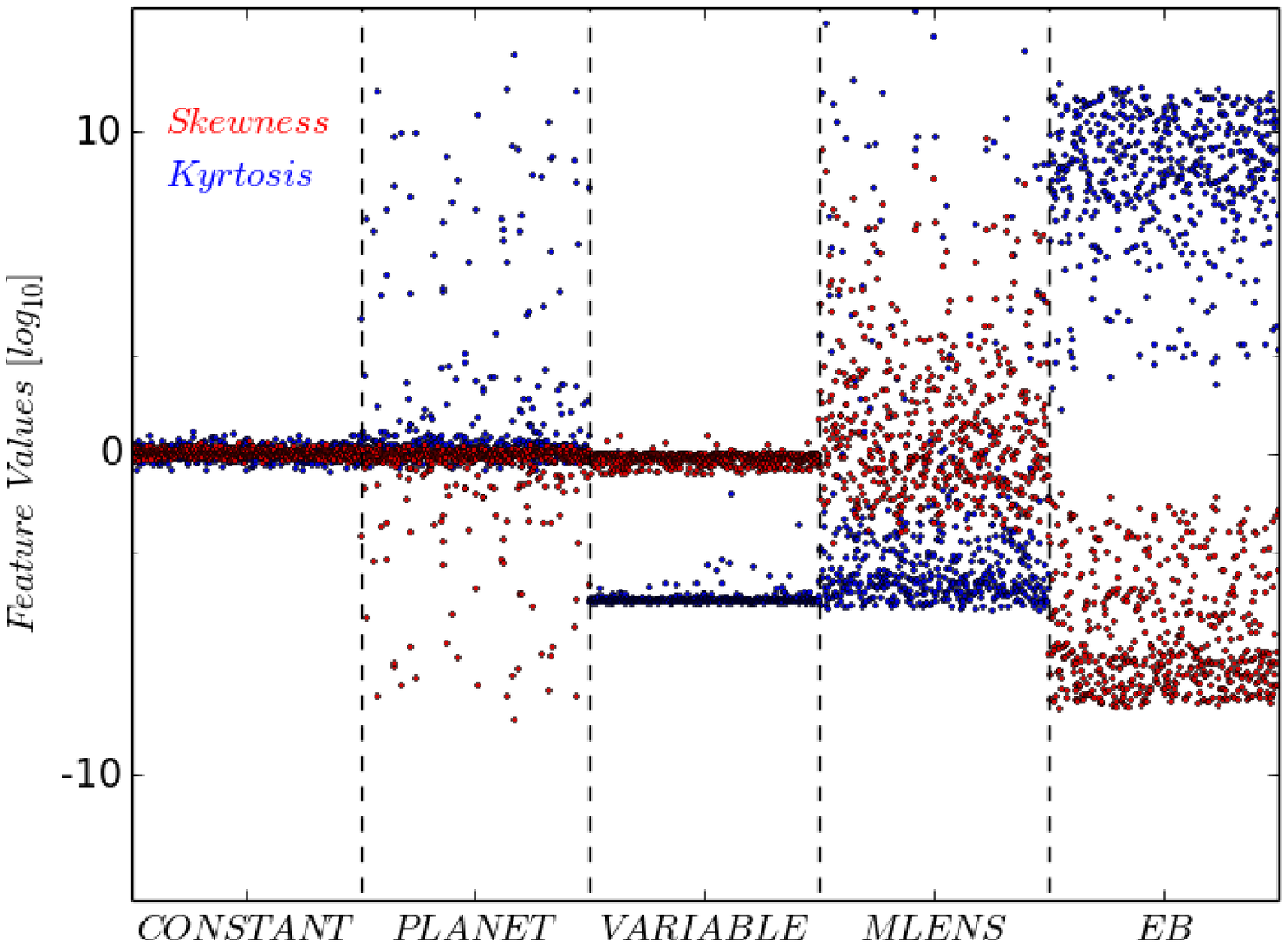} \\
\includegraphics[width=9cm]{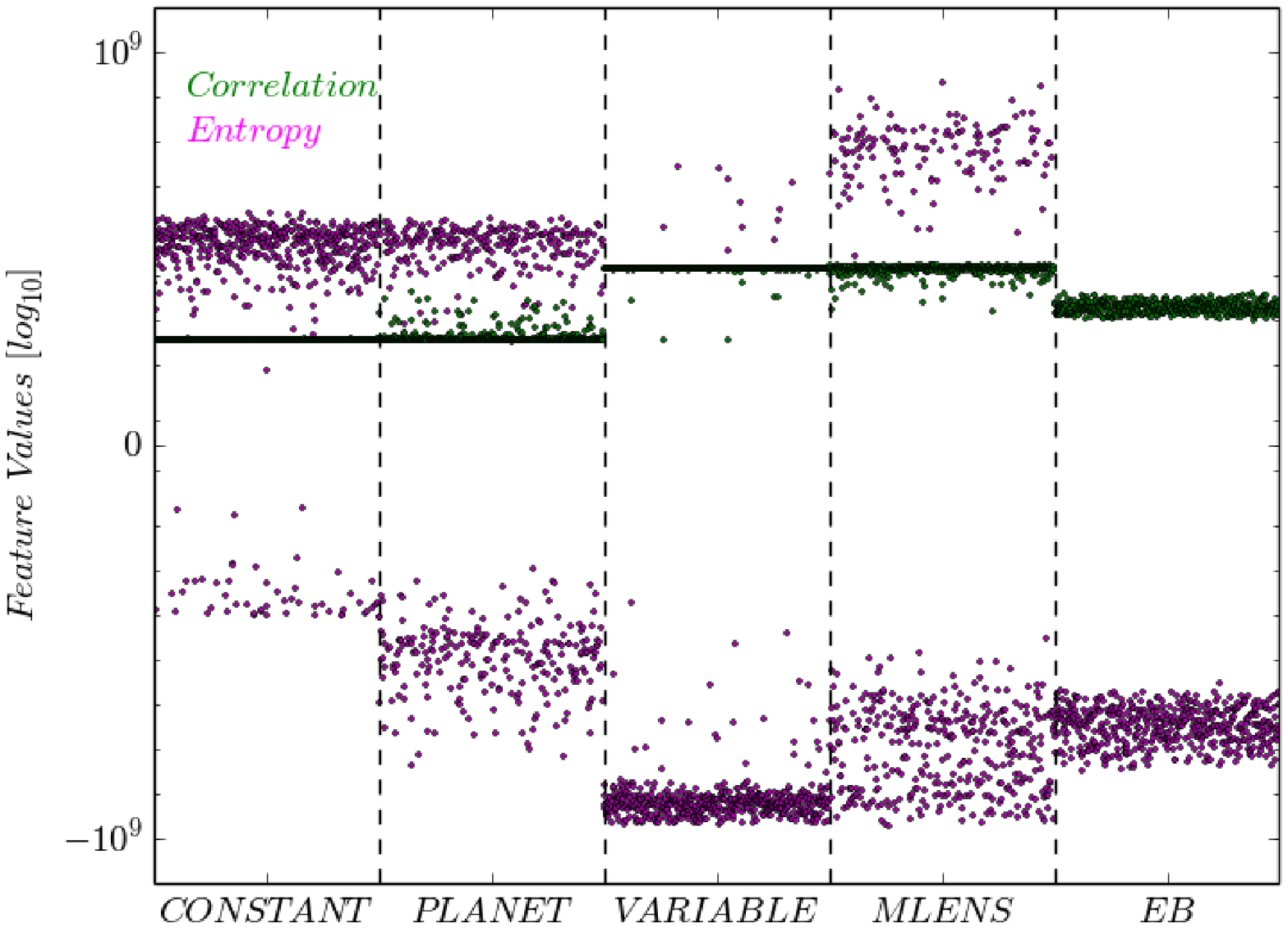}
\caption{Different features for all classes. Top: the skewness and kurtosis (red and blue, respectively). Bottom: autocorrelation and entropy features (green and magenta, respectively).}
\label{fig4}
\end{figure}

\section{Performance } \label{sec:perf}
\subsection{General} \label{sec:general}

\begin{figure}
\centering
\includegraphics[width=10cm]{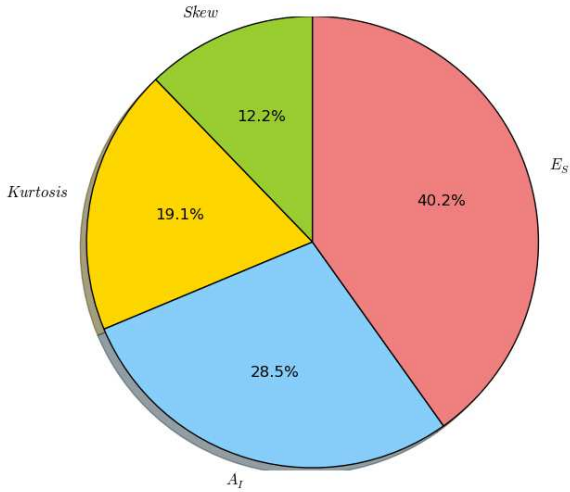}
\caption{The feature importance statistics. $E_{\rm S}$ value is the most important (40.2\%).}
\label{fig2}
\end{figure}

First we create a 'TRAIN' sample using the five classes of light curves described in Section \ref{sec:lcsim}. We use 1000 light curves for each class (5000 in total) and we 
calculate the $S$, $K$, $A_{\rm I}$ and $E_{\rm S}$ features of each one. We also add a flag (CONSTANT, TRANSIT, VARIABLE, EB, MLENS) for each algorithm decision per class. By 
training \texttt{SIDRA} we found that the fitting score of the `TRAIN' sample is 90\% using 100 trees. That means from the 5,000 light curves, \texttt{SIDRA} could successfully 
distinguish 4,500 of them. The majority of the remaining 10\%, which \texttt{SIDRA} fails, comes from CONSTANT and TRANSIT due to low S/N ratio, as we have described 
previously. \\
It is very important for all features to have a good statistical weight in the procedure. The importance of each feature is high enough to be included in the algorithm. After the 
training procedure, the importance of each feature is shown in Fig. \ref{fig2}.
\noindent
The most important feature is $E_{\rm S}$ with 40.2\%, then $A_{\rm I}$ with 28.5\% and skewness and kurtosis with 12.2 and 19.1\%, respectively. The $A_{\rm I}$  feature contains 
high values for high\textendash amplitude light curves such as microlensing and/or variables. On the other hand, skewness and kurtosis include information on the light curve shape, 
which is different for different classes. Finally, $E_{\rm S}$ shows values around zero for variables, and constant, with high positive values for microlensing and negative values 
for planets and binaries. \\
We applied \texttt{SIDRA} to 10000 light curves from all classes as a blind test (50000 light curves in total). We create a \textit{confusion matrix} using these 
results. In principal, the algorithm collects all decisions from all different trees. The final decision is made by maximizing the probability from all of the 
different trees. In the worst case, the minimum decision probability is 0.2 (because we have five classes\textendash 1/5), for a five class Random Forest such as \texttt{SIDRA}.\\ 
\begin{figure}
\centering
\includegraphics[width=9.5cm]{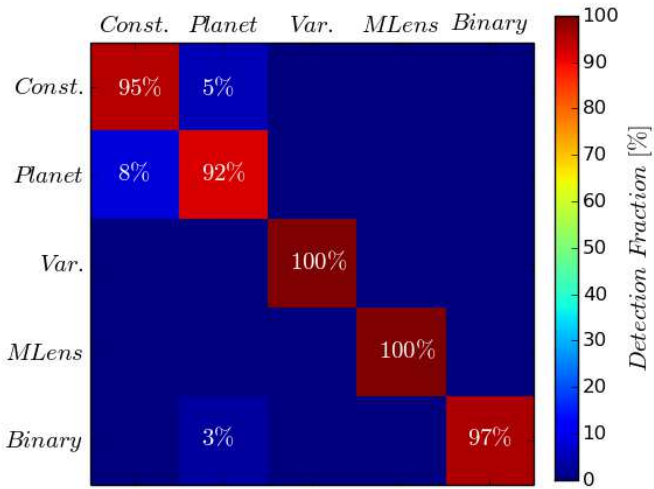}
\caption{The confusion matrix for decision probability of 0.6, where rows and columns refer to input and output light\textendash curve flag, respectively. The color\textendash bar 
refers to the success ratio.}
\label{fig3}
\end{figure}
\noindent
It is obvious that the small decision probability decreases the success ratio of the algorithm because we have to deal with a flip-coin decision. We force \texttt{SIDRA} to 
take more certain decisions. For our example, we used a decision probability equal to 0.6. The \textit{confusion matrix} shows the results of this test (Fig. \ref{fig3}). The success 
ratio for each class is 100\% for microlensing and variable stars, 97\% for eclipsing binaries (3\% planet false alarm), 95\% constant (5\% planet false alarm) and 92\% for 
transits (8\% constant false alarms). If we increase the decision probability (from 0.2 0.6) some light curves in the range of 0.2\textendash 0.6 are rejected. At 0.6 decision probability 
the algorithm rejects 5\% of the total sample. \\
It is clear that the success ratio of the algorithm is a function of the decision probability cut and there are no `golden' fixed numbers for each survey. On the other hand, each 
survey should define these numbers for their own goals, depending on their targets and features. As an example we can say that it is extremely rare for a transit survey to detect 
a microlensing event. Most of the transit surveys (if not all) avoid the fields in the Galactic plane. In these fields the probability to detect a microlensing event is close 
to zero. On the other hand microlensing surveys do not search for transiting planets because of the magnitude range and faint target stars of the field. \\
We plan to give a more detailed analysis for the decision probability based on different algorithms (such as Bayesian, Dempster\textendash Shafer theory and/or Fuzzy Logic) in a 
future paper.

\subsection{A closer look at planets} \label{sec:plan}
\begin{figure}
\centering
\includegraphics[width=9.5cm]{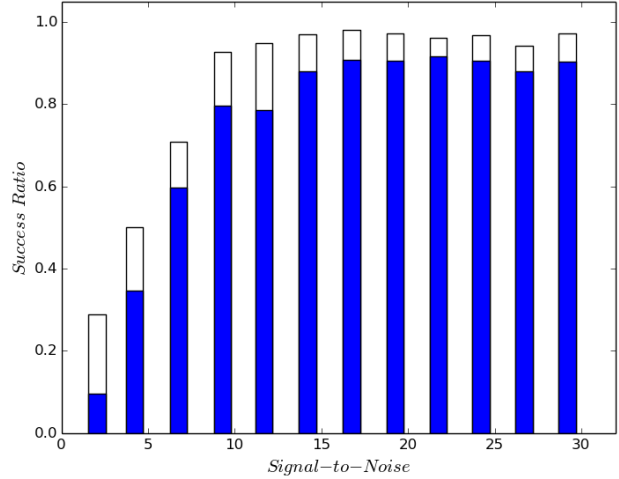} 
\caption{BLS (blue bars) and \texttt{SIDRA} (white bars) results after simultaneously searching for different S/N values.}
\label{fig6}
\end{figure}

From the tests in Section \ref{sec:general}, the most confusing classes are CONSTANT and TRANSIT. Fig. \ref{fig3} suggests that 92\% of the transiting light curves can be resolved 
by \texttt{SIDRA}, but this is not totally true. In Table \ref{tab:t1} we select all the host stars between F0 and M5, planetary radius 0.7\textendash 2$R_{\rm J}$ and noise from 
1\textendash 5\%. That means that for 30\% of our stellar sample (F0\textendash G0), the transit depth range is from 0.008 to 2.5\% for the worst and best case respectively. 
Most of these planets do not show any signal in the light curve with the noise properties we used and it is impossible to be detected. The real question is how many `detectable' 
transiting planets \texttt{SIDRA} could flag. \\
In order to judge the algorithm on the transiting light curves sample, we compare \texttt{SIDRA} with a BLS\textendash type algorithm, such as \cite{kovacs}. We run BLS and 
\texttt{SIDRA} simultaneously on the same data set. For BLS we select a signal detection threshold at the $1\sigma$ level. Even if $1\sigma$ is not realistic for a real world 
survey (we expect signals above 2$\sigma$), this threshold is generous for BLS. If we increase the detection threshold, of course we expect much less planets. We compare with 
\texttt{SIDRA} 0.5 decision probability threshold (Fig \ref{fig6}).  
\noindent
Both of the algorithms found approximately the same amount of planets. \texttt{SIDRA} detected 85.4\% of the sample and BLS 77.9\% of the sample (7.5\% less planets than 
\texttt{SIDRA}) even with 1$\sigma$ detection threshold. Also, Fig. \ref{fig6} shows that \texttt{SIDRA} is more sensitive than BLS for low S/N light curves.

\subsection{Real data example - Kepler Mission} \label{sec:kepler}

\begin{figure}
\centering
\includegraphics[width=9.5cm]{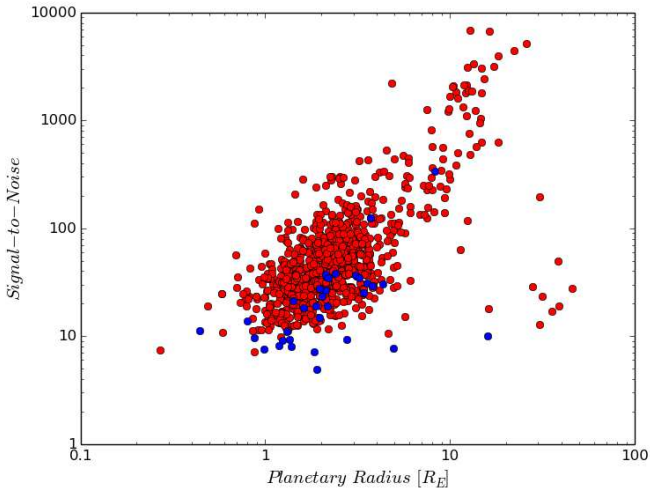}
\caption{The \textit{Kepler} sample we used for our test. From 0.5 to 50 $R_{\rm E}$ and 5\textendash 5000 S/N. The $S/N$ has been calculated in a 400d orbital period window. 
Plot shows also the \texttt{SIDRA} successful and unsuccessful detections (red and blue dots respectively).}
\label{kep_sam}
\end{figure}

\begin{figure}
\centering
\includegraphics[width=9.5cm]{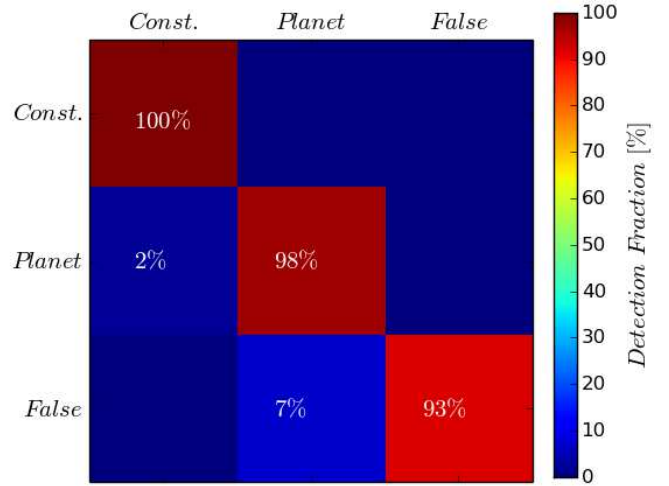}
\caption{The confusion matrix of the \textit{Kepler} data test, where rows and columns refer to input and output light\textendash curve flag, respectively. The color\textendash bar 
refers to the success ratio. Decision probability is 0.5.}
\label{kep}
\end{figure}

The next task was to use our algorithm on real data in order to check its success. This section is only a small example with limited amount of data and tests. We plan to present a 
full \textit{Kepler} mission data analysis using \texttt{SIDRA} in a future publication. For our tests we focus only on the exoplanets group of light curves using \textit{Kepler} 
public light curves available from the \textit{Kepler} archive hosted by the Multi\textendash mission Archive at STScI\footnote{http://archive.stsci.edu}. The observations comprised 
only from the long-cadence.\\
In order to use these data from a space mission, we modified our training sample. We did not train for variables, microlensing events or eclipsing binaries. \\
We have used \textit{Kepler} Q1\textendash Q6 KOI data set. We used 2000 light curves flagged as transit candidates (PLANET-SET) \citep{batalha, mullally} and 2000 light curves 
from the Q1\textendash Q6 data set flagged as non-transiting exoplanets (CONSTANT-SET). We select our sample randomly, which means that our sample includes both large and small 
transiting light curves. Fig \ref{kep_sam} shows a planetary radius versus S/N ratio of the \textit{Kepler} sample we used.  
\noindent
Furthermore we create another data set with 1000 light curves flagged as FALSE-SET. These light curves are included in the KOI \textit{Kepler} catalogue but they are not real 
planets. We select false alarm light curves which belong to the CONSTANT-SET. Finally, we have to mention that we were using 400d of Q1\textendash Q6 KOI data set.\\
We train the forest using two classes (CONSTANT and TRANSIT) as described in Section 2.1, but we run it for all three data sets. We used rms range from 0.001 to 0.01, period range 
1\textendash 200d and planetary radius of 1\textendash 5 $R_{\rm E}$. The total number of training light curves was 1000 per class. Fig. \ref{kep} shows the results of our test. 
\noindent
The success ratio of real planets is 98\%. The constant success ratio is 100\% and the False Alarm success ratio is 93\%. The False Alarm ratio is quite important. \texttt{SIDRA} 
classifies only the 7\% of the False Alarm light\textendash curves as planets, which appear in the KOI \textit{Kepler} list. On the other hand \texttt{SIDRA} seems to 
miss 2\% of real planets. In order to detect planets with higher period and/or smaller S/N, we need many more data than 400d. Also, we did not include any `exotic' light 
curves in our training sample (Section \ref{sec:exotic}). A more accurate analysis is required in a future paper including many more classes other than Constant and Planet data sets. 
The decision probability of 0.5 contains the 90\% of the sample.

\subsection{Exotic light curves} \label{sec:exotic}
\begin{figure}
\centering
\includegraphics[width=9.5cm]{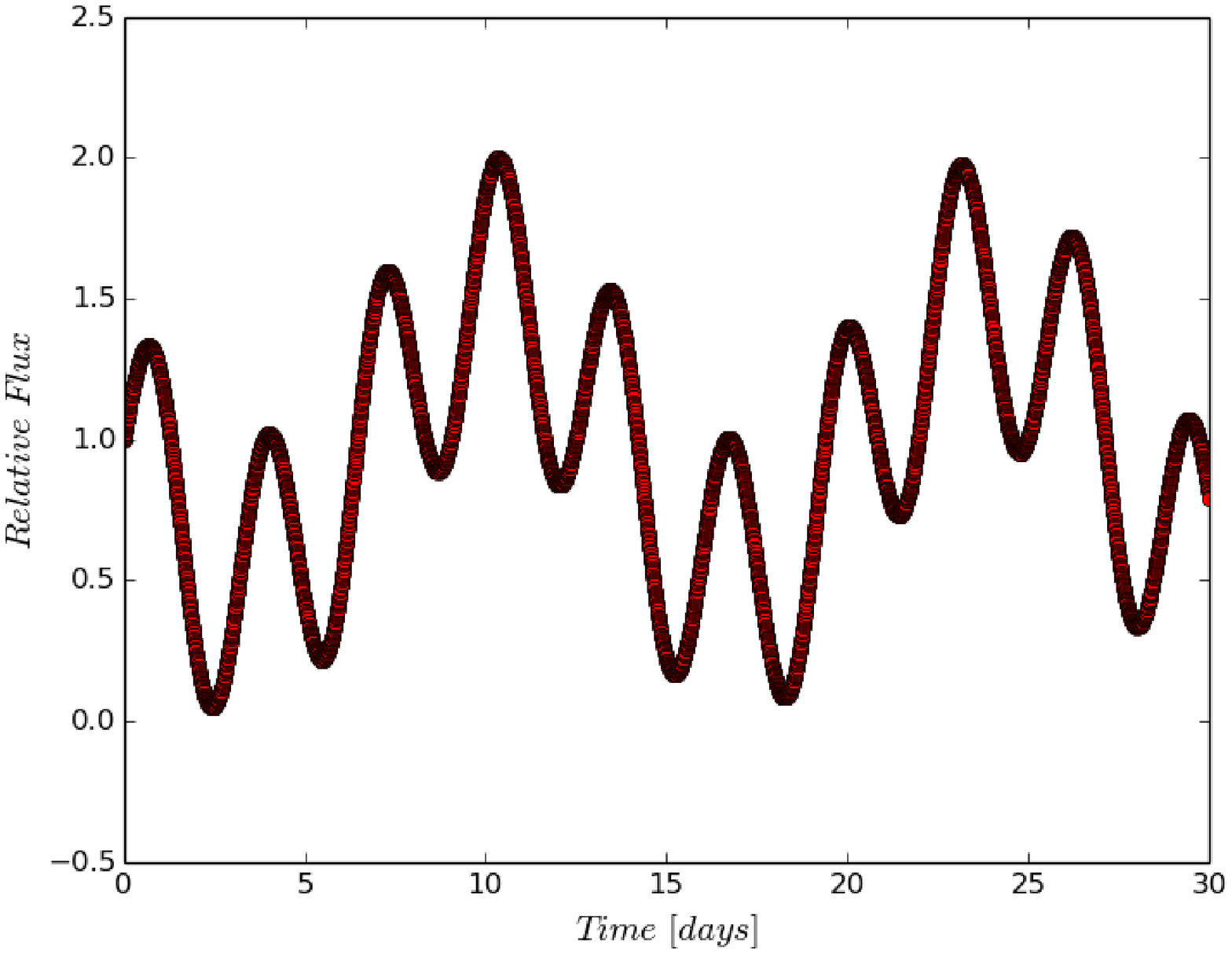} \\
\includegraphics[width=9.5cm]{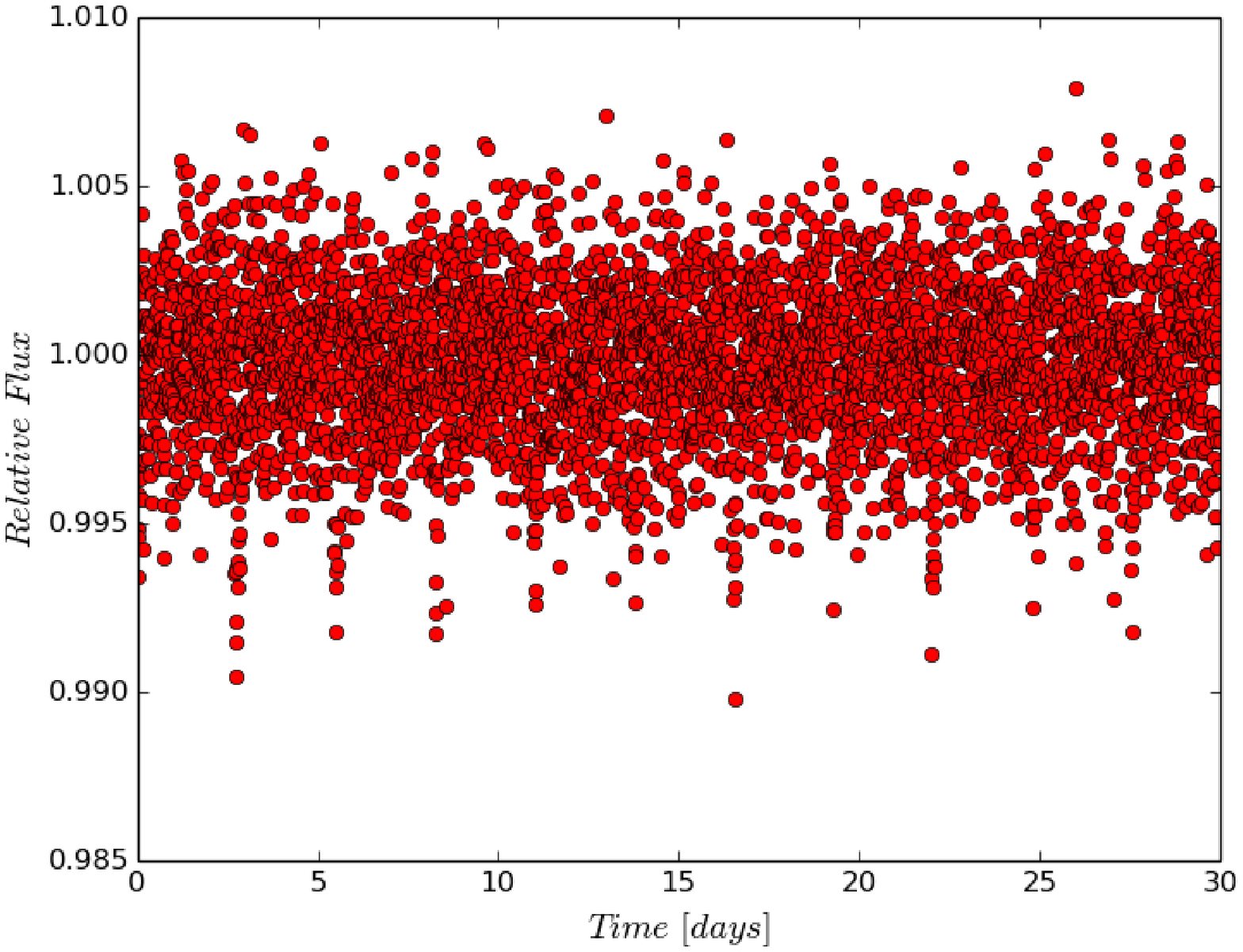}
\caption{Top: a 30d multi\textendash period variable star. Bottom: the same light curve after removing strong periodicities.}
\label{multi}
\end{figure}

The \textit{Kepler} mission data have shown how difficult it is to detect transiting exoplanets around a star with high variability. These kind of light curves are the most important for space 
missions because with such high photometric accuracy, most of the stars show some kind of real variability. BLS-like algorithms fail to detect these kind of transits because of the 
algorithm design. BLS assumes that the out-of-transit mean value of all transits is 1 (or 0). This is not true in a variable star light cure with transit. The signal 
of the variable star dominates the light curve with very strong primary and harmonic periods. Almost all the combined transiting light curves need special analysis. For a pure blind 
detection method it is a very difficult problem for any algorithm, including Machine Learning.\\
\texttt{SIDRA} is able to `solve' the problem with a combined analysis method similar to other BLS-like techniques. We simulate multi\textendash period variable plus transiting 
exoplanet signals using \textit{Kepler} accuracy specifications. Fig. \ref{multi}\textendash (top) shows an example of our simulated light curve.
\noindent
The strategy is simple. We first run \texttt{SIDRA} on the raw light curve. The algorithm classifies the light curve as a variable with probability 98\%. Once we detect a variable 
signal, we use Lomb\textendash Scargle, FFT or binned polynomial fits in order to remove strong periodicities (Fig. \ref{multi}\textendash Bottom). Finally, we run \texttt{SIDRA} 
once more using the new light curve. We detect a planet with probability of 81\%. \\
We run the same experiment using BLS and it was able to detect the planet. We do not claim of course that this technique is new or it does not work with other detection algorithms. 
We just give an example showing that \texttt{SIDRA} is also able to detect exoplanets hidden in a strong variability.

\section{Conclusions} \label{sec:discussion}

\texttt{SIDRA} is a blind detection and classification algorithm based on Machine Learning\textendash Random Forest technique. This paper is a general presentation of the algorithm. 
We used simulated light curves from five different classes. These are constant, transiting, variables, microlensing and eclipsing binary light curves. Assuming a 60\% decision 
probability, the algorithm success ratio is 95-100\% for microlensing, eclipsing binaries, variables and constant light curves and 91\% for transits for Table \ref{tab:t1} 
input values. Also we test \texttt{SIDRA} with real light curves from the \textit{Kepler} mission. We detect and classify successfully 100\%, 98\% and 93\% of the constant, 
transiting and false alarm light curves. Furthermore, we show a simulated example of \texttt{SIDRA} transit detection around a variable host star.\\
We discuss the transiting exoplanet detection power compared with BLS-like algorithms, but we would like to make clear that we do not suggest to replace BLS with \texttt{SIDRA}, 
even if in our test \texttt{SIDRA} detects 8\% more planets and is 1000 times faster. What we suggest for a at least an exoplanet survey, is to include both algorithms. 
\texttt{SIDRA} could be a very powerful tool, and could easily detect/classify interesting objects which require further analysis. 

\subsection{Advantages} \label{sec:adva}

The algorithm could be easily modified for each team and project and it is as general as possible, solving simultaneously different types of light curves. On the other hand, BLS or 
high $\Delta \chi^{2}$ methods for example, work only for transiting and eclipsing binary light curves. \\
For eclipsing binaries, variables and microlensing light curves, the algorithm is not only detecting the signal correctly but it also minimizes the false alarms. In our case of 
10000 simulated light curves, false alarms for these objects were eliminated. \\
For transiting light curves, it manages to detect more planets than the classical method of BLS. Furthermore, \texttt{SIDRA} is much faster than BLS. Once we train the network, the 
classification needs 4 ms for a single light curve of 4300 data points, making \texttt{SIDRA} ideal for huge surveys such as \textit{TESS} and \textit{PLATO} transiting exoplanet 
future space missions. BLS needs $\sim$4 s in the same machine for the same light curve. \\
Finally, \texttt{SIDRA} has the ability to become a huge network with almost all kinds of light curves. We can not only classify variable stars for example, but we can use many more 
classes in order to identify the type of each variable. On the other hand, we are able to search for non-periodic phenomena such as supernova and flare stars. 

\subsection{Disadvantages} \label{sec:disadva}

The major disadvantage of the algorithm is that it cannot resolve physical characteristics from the light curves because of its nature. For example we cannot extract the information 
about the radius of the planet because we do not fit physical parameters but we calculate statistical values of each light curve. Of course for the periodic events, the information 
of the period is hidden in the autocorrelation function, but still there is more information which is missing.\\
Finally, the algorithm works with a training light curve set, which means that we have to be very careful on the selection of features and limits in order to maximize the success 
of the algorithm. 

\subsection{False Alarms} \label{sec:alarms}

The main problem of every detection and classification algorithm is the false positive and negative alarms. Assuming a 20000 light curve sample, we expect from \texttt{SIDRA} to 
solve the majority of the light curves correctly. On the other hand there are some limitations. We assume that the 1\% of the theoretical sample contains real planets, 15\% real 
variables, 10\% real eclipsing binaries and 0\% real microlensing events. Table \ref{tab:t2} shows the results. From the total number the 95\% remain in our sample after the 
decision probability 0.6 cut. These remaining light curves (5\%) are flagged as unknown.\\ 
In order to deal with the 4\% of false alarm and the 5\% of unknown light curves, we can decrease step-by-step the decision probability from 0.6. This will increase the 
\texttt{SIDRA} sample, decreasing the unknown light curves. Also, we can use a typical BLS algorithm. It is not clear that BLS could solve the false alarms better than \texttt{SIDRA} 
(Fig. \ref{fig6}). That depends on the S/N of each light curve but we can use BLS as a separate tool. As we mention above, the decision probability is a very important variable 
and we plan to study it in a future publication.  
\begin{table*}
\centering
\caption{\label{tab:t2} Classes statistics assuming a 20000 light curve sample.}
\begin{tabular}{lccccc}
Classes & Total number & \texttt{SIDRA} sample $(>0.6)$ & Successful detection & False alarms & Unknown    \\
\hline
Constant & 14800 & 14060 & 13357 & 703 as planets & 740   \\
Planet & 200 & 190 & 175 & 15 as constant& 10   \\
Variable & 3000 & 2850 & 2850 &  & 150   \\
EB & 2000 & 1900& 1843& 57 as planets& 100   \\
\hline

Total & 20000 & 19000 (95\%)& 18225 (91\%)& 775 ($\sim$ 4\%) & 1000 (5\%)   \\

\hline
\end{tabular}
\end{table*}

\section*{Acknowledgements}

This publication is supported by NPRP grant no. X-019-1-006 from the Qatar National Research Fund (a member of Qatar Foundation). The statements made herein are solely the 
responsibility of the authors.

\bsp

\label{lastpage}


\begin{thebibliography}{99}

\bibitem[\protect\citeauthoryear{Alcock et al.}{2000}]{alcol} Alcock C. et al. 2000, ApJ, 541, 734
\bibitem[\protect\citeauthoryear{Alsubai et al.}{2013}]{alsubai} Alsubai K. et al. 2013, Acta Astron., 63, 465
\bibitem[\protect\citeauthoryear{Bakos et al.}{2011}]{bakos} Bakos G., Hartman J., Torres G., Kov\`{a}cs G., Noyes R. W., Latham D. W., Sasselov D. D.; B\'{e}ky B., 2011, 
EPJ Web. Conf., 11, 01002
\bibitem[\protect\citeauthoryear{Batalha et al.}{2013}]{batalha} Batalha N. et al. 2013, ApJS, 204, 24
\bibitem[\protect\citeauthoryear{Bond et al.}{2001}]{bond} Bond I. et al. 2001, MNRAS, 327, 868
\bibitem[\protect\citeauthoryear{Bonomo et al.}{2012}]{bonomo} Bonomo A. S. et al. 2012, A\&A, 547.110
\bibitem[\protect\citeauthoryear{Bord\'{e} et al.}{2007}]{borde} Bord\'{e} P., Fressin F., Ollivier M., L\'{e}ger A., Rouan D., 2007, in Afonso C., Weldrake D.,
Henning Th., eds, ASP Conf. Ser.- Vol. 366, Transiting Extrasolar Planets Workshop. Astron. Soc. Pac., San Francisco, p. 145
\bibitem[\protect\citeauthoryear{Boruki et al.}{2010}]{boruki} Borucki W. et al. 2010, A\&AS, 21510101
\bibitem[\protect\citeauthoryear{Breiman}{2001}]{Breiman} Breiman L., 2001, Mach. Learn., 45, 5
\bibitem[\protect\citeauthoryear{Cabrera et al.}{2012}]{cabrera} Cabrera J., Csizmadia Sz., Erikson A., Rauer H., Kirste S., 2012, A\&A, 548, 44
\bibitem[\protect\citeauthoryear{Cameron et al.}{2006}]{cameron} Cameron C. A. et al. 2006, MNRAS, 373, 799
\bibitem[\protect\citeauthoryear{Claret}{2004}]{claret} Claret A., 2004, A\&A, 428, 1001
\bibitem[\protect\citeauthoryear{Foreman-Mackey et al.}{2015}]{Foreman-Mackey} Foreman-Mackey D., Montet B., Hogg D., Morton T. D., Wang D., Sch\"{o}lkopf B., 2015, ApJ, 806, 13
\bibitem[\protect\citeauthoryear{Hogg et al.}{2013}]{KeplerML} 	Hogg D.W. et al., 2013, Kepler Project Office Call for White Papers: Soliciting Community 
Input for Alternate Science Investigations for the Kepler Spacecraft
\bibitem[\protect\citeauthoryear{Kaler}{1998}]{kaler} Kaler J. B., 1998, Mon. Notes Astron. Soc. South. Afr., 57, 89
\bibitem[\protect\citeauthoryear{Kov\'{a}cs et al.}{2002}]{kovacs} Kov\'{a}cs G., Zucker S. \& Mazeh T., 2002, A\&A, 391, 369
\bibitem[\protect\citeauthoryear{Li et al.}{2008}]{Li} Li L., Zhang Y., \& Zhao Y., 2008, Sci. China G, 51, 916
\bibitem[\protect\citeauthoryear{McCauliff et al.}{2014}]{mccauliff} McCauliff S., Jenkins J., Catanzarite J., 2014, A\&AS, 22412009
\bibitem[\protect\citeauthoryear{Masci et al.}{2014}]{masci} Masci F. J., Hoffman D. I., Grillmair C. J., Cutri R. M., 2014, AJ, 148, 21
\bibitem[\protect\citeauthoryear{Moutou et al.}{2007}]{moutou} Moutou C. et al. 2007, in Afonso C., Weldrake D., Henning Th., eds, ASP Conf. Ser.-Vol. 366, Transiting Extrasolar 
Planets Workshop. Astron. Soc. Pac., San Francisco, p. 127
\bibitem[\protect\citeauthoryear{Mullally et al.}{2015}]{mullally} Mullally F. et al. 2015, ApJS, 217, 31
\bibitem[\protect\citeauthoryear{Paczynski}{1986}]{paczynski} Paczynski B., 1986, ApJ, 304, 1
\bibitem[\protect\citeauthoryear{P\'{a}l et al.}{2008}]{pal} P\'{a}l A., 2008, MNRAS, 390, 281
\bibitem[\protect\citeauthoryear{Pr\v{s}a et al.}{2005}]{prsa} Pr\v{s}a A. \& Zwitter T., 2005, ApJ, 628, 426
\bibitem[\protect\citeauthoryear{Robin et al.}{2004}]{robin} Robin A. C., Reyl\'{e} C., Derri\`{e}re S., Picaud S., 2004, A\&A, 416, 157
\bibitem[\protect\citeauthoryear{Shannon et al.}{1949}]{shannon} Shannon C., Weaver W., 1949, The Mathematical Theory of Communication. Univ. Illinois Press, Urbanamtc
\bibitem[\protect\citeauthoryear{Sumi}{2011}]{sumi} Sumi T., MOA \& OGLE Collaboration, 2011, ESS.2.0103
\bibitem[\protect\citeauthoryear{Udalski et al.}{1992}]{udalski} Udalski A., Szymanski M., Kaluzny J., Kubiak M., Mateo M., 1992, Acta Astron., 42, 253
\bibitem[\protect\citeauthoryear{Wyrzykowski et al.}{2014}]{wyrzykowski} Wyrzykowski L. et al. 2014, Acta Astron., 64, 197


	
	
\end{thebibliography}
\end{document}